\documentclass[aps,prb,twocolumn,floatfix,showpacs]{revtex4}
\usepackage{amsmath}
\usepackage{amssymb}
\usepackage{graphicx}
\usepackage{bm}

\newcommand{\eref}[1]{eqn.~(\ref{#1})}

\newcommand{\erefs}[2]{eqns.~(\ref{#1},\ref{#2})}

\newcommand{\fref}[1]{fig.~\ref{#1}}
\newcommand{\Fref}[1]{Fig.~\ref{#1}}

\newcommand{\sref}[1]{sec.~\ref{#1}}
\newcommand{\srefs}[2]{secs.~\ref{#1},\ref{#2}}

\newcommand{\srefseq}[2]{secs.~\ref{#1} - \ref{#2}}

\DeclareMathOperator{\Tr}{Tr}
\renewcommand{\Im}{\mathrm{Im}\,}
\newcommand{\mub}{\mu_\mathrm{B}}

\newcommand{\pd}{\phantom{\dagger}}
\newcommand{\Hdot}{\hat{H}_{\mathrm{D}}}
\newcommand{\Hl}{\hat{H}_{\mathrm{L}}}
\newcommand{\Ht}{\hat{H}_{\mathrm{T}}}

\newcommand{\half}{\tfrac{1}{2}}
\newcommand{\K}{\mathbf{k}}

\newcommand{\tk}{T_\mathrm{K}}
\newcommand{\tkn}{T_\mathrm{K}^0}

\newcommand{\ek}{\epsilon_{\mathbf{k}}^{\pd}}
\newcommand{\acre}{a^{\dagger}_{\mathbf{k} \nu \sigma}}

\newcommand{\ades}{a^{\phantom\dagger}_{\mathbf{k} \nu \sigma}}
\newcommand{\adesL}{a^{\phantom\dagger}_{\mathbf{k} L \sigma}}
\newcommand{\adesR}{a^{\phantom\dagger}_{\mathbf{k} R \sigma}}
\newcommand{\ccre}{c^{\dagger}_{\mathbf{k}\sigma}}
\newcommand{\cdes}{c^{\phantom\dagger}_{\mathbf{k} \sigma}}
\newcommand{\dicre}{d^{\dagger}_{i\sigma}}
\newcommand{\dides}{d^{\phantom\dagger}_{i\sigma}}

\newcommand{\dcreo}{d^{\dagger}_{1\sigma}}
\newcommand{\ddeso}{d^{\phantom\dagger}_{1\sigma}}
\newcommand{\nis}{\hat{n}_{i\sigma}^{\pd}}
\newcommand{\nims}{\hat{n}_{i-\sigma}^{\pd}}
\newcommand{\nic}{\hat{n}_{i}^{\pd}}

\newcommand{\nso}{\hat{n}_{1\sigma}^{\pd}}
\newcommand{\nmso}{\hat{n}_{1-\sigma}^{\pd}}

\newcommand{\none}{\hat{n}_{1}^{\pd}}
\newcommand{\ntwo}{\hat{n}_{2}^{\pd}}

\newcommand{\up}{U^{\prime}}

\newcommand{\shat}[1]{\hat{\textbf{s}}_{#1}^{\pd}}

\newcommand{\jtil}{\tilde{J}_{\mathrm{H}}}
\newcommand{\util}{\tilde{U}}
\newcommand{\uptil}{\tilde{\up}}

\newcommand{\etil}{\tilde{\epsilon}}
\newcommand{\htil}{\tilde{h}}

\newcommand{\expn}[1]{\langle\hat{n}_{#1}\rangle}

\newcommand{\Simp}{S_{\mathrm{imp}}}

\newcommand{\Tkone}{T_K^{S=1}}

\newcommand{\expnum}[2]{#1\times 10^{#2}}

\begin{document}

\title{Magnetic field effects in few-level quantum dots: theory, and application to experiment}

\date{July 12, 2011}
\author{Christopher J. Wright}
\author{Martin R. Galpin}
\author{David E. Logan}
\affiliation{Oxford University, Chemistry Department, Physical \& Theoretical Chemistry, 
South Parks Road, Oxford, OX1~3QZ, UK.}

\begin{abstract}
We examine several effects of an applied magnetic field on Anderson-type models for both single- and two-level quantum dots, and make direct comparison between numerical renormalization group (NRG) calculations and recent 
 conductance measurements. On the theoretical side the focus is on magnetization, single-particle dynamics and zero-bias conductance, with emphasis on the universality arising in strongly correlated regimes; including 
a method to obtain the scaling behavior of field-induced Kondo resonance shifts over a very wide field range. NRG is
also used to interpret recent experiments on spin-$\half$ and spin-$1$ quantum dots in a magnetic field, which we argue do not wholly probe universal regimes of behavior; and the calculations are shown to yield good qualitative 
agreement with essentially all features seen in experiment. The results capture in particular the observed field-dependence of the Kondo conductance peak in a spin-$\half$ dot, with quantitative deviations from experiment occurring at fields in excess of $\sim 5 \mathrm{T}$, indicating the eventual inadequacy of using the equilibrium single-particle spectrum to calculate the conductance at finite bias.
\end{abstract}


\maketitle

\section{Introduction}
\label{sec:intro}

Understanding electronic transport in quantum dots remains a major challenge for theorists working on correlated electron systems. Conductance at low energies is often dominated by one of a number of Kondo effects,\cite{Hewson1993} in which the strong localized interactions on the dot(s) induce non-trivial many-body physics. Over the years a wide range of such Kondo effects have been predicted and observed, in single and multiple quantum dot devices of various geometries.\cite{Kouwen1997,Wiel2003,Pustil2004,Anderg2010,Ilani2010}

Here we consider a single quantum dot, tunnel-coupled to two leads\cite{Glazma1988,Ng1988,Meir1992,Goldha1998,Cronen1998} in an effective one-channel fashion. While 
the dot will in general hold many electrons in its quantized levels, only those close to the Fermi level contribute in practice to electronic transport provided the mean level spacing is sufficiently large, and the rest can be 
neglected with relative impunity. Typically just one level is important, but occasionally one observes the case of two relevant levels, where the physics is richer; including e.g. a
quantum phase transition between Fermi liquid and underscreened Kondo phases\cite{Pustil2000,Kikoin2001,Hofste2001,Pustil2001,Vojta2002,Pustil2003,Hofste2004,Pustil2006,Zitko2006,Zitko2007,Posazh2007,Roura2009,Logan2009,Cornag2011,Florens2011} which has been observed in several experimental guises.\cite{Kogan2003,Roch2009,Parks2010}

We present and examine critically a number of results falling under the umbrella of magnetic field ($B$) effects in these single- and two-level quantum dots, the appropriate models for which are specified in \sref{sec:models}.
The paper consists of two related parts. In the first (\srefs{sec:zerofield}{sec:statdynam}), using mainly Wilson's numerical renormalization group (NRG) method,\cite{Wilson1975,Krishn1980,Bulla2008} we consider magnetization, single-particle dynamics and the zero-bias conductance, with emphasis on the universality and scaling behavior arising in the strongly correlated regimes of the models.

 Even for single-level quantum dots described by an Anderson impurity model,\cite{Anders1961,Hewson1993} 
there are still open questions regarding single-particle dynamics in the presence of a magnetic field;
our primary concern being the field-induced Kondo peak splitting in the equilibrium 
single-particle  spectrum. This has been analyzed by a number of authors and techniques,\cite{Costi2000,Hofste2000,Moore2000,Logan2001,Hewson2006,Bauer2007,Bauer2007a,Zitko2009,Zitko2011}  but the results are not in complete agreement.\cite{Costi2000,Moore2000,Logan2001,Quay2007} We show that there exists an algorithm by which NRG can obtain the universal behavior over many orders of magnitude of field strength but that, eventually, even the most accurate NRG calculations cannot completely resolve the universal splitting at 
very  large fields. The corresponding situation for the two-level model is considered in \sref{ssec:dynam}.
We also obtain the field and temperature ($T$) dependence of the zero-bias conductance, and for $T=0$ in particular 
generalize the Luttinger integral analysis of ref.~\onlinecite{Logan2009} to encompass a finite magnetic field; leading to an exact result for the conductance for any field, and insight into the rather subtle differences between the limits $B=0$ and $B \rightarrow 0$ for the underscreened triplet phase of the model.

In the second part of the paper (\sref{sec:experiment}) we turn to comparison with experiment. Two recent sets of conductance measurements on quantum dots in a magnetic field are considered, from the groups of Kogan\cite{Liu2009} (on an effective one-level dot) and Goldhaber-Gordon\cite{Quay2007} (on both effective one- and two-level systems). From comparison to NRG results, we are able to determine reliable bare model parameters for the Anderson-type (as opposed to Kondo) models considered in \srefseq{sec:models}{sec:statdynam}, as relevant to experiment. With these, our NRG calculations are shown to yield very good qualitative agreement with essentially all features observed in both experiments.\cite{Liu2009,Quay2007}
In particular, we show that theory can in fact explain the evolution of the field-induced
splitting of the Kondo conductance peak observed in ref.~\onlinecite{Liu2009} --
including a simple explanation for an observed crossing in the peak splittings of two different quantum dots. The agreement is essentially quantitative up to field strengths of around a couple of Kondo scales, but beyond that our calculations deviate from the experimental data.
This reinforces results from a recent study using the scattering states NRG\cite{Schmitt2011} and earlier renormalized perturbation theory and NRG calculations,\cite{Hewson2005,Hewson2006} showing that the commonly used approximation of calculating the source-drain bias dependence of the conductance from the equilibrium spectrum is unsuitable for making quantitative comparisons to experiment sufficiently far out of equilibrium. Indeed, until more progress in non-equilibrium theory is made, we suggest that experiments should instead aim to make comparison with the magnetic-field dependence of the zero-bias conductance.

\section{Models}
\label{sec:models}
Each model considered in this work consists of a single interacting quantum dot region, tunnel coupled to a pair of non-interacting metallic leads. As mentioned above, we focus on the situation where the mean level spacing of the dot is sufficiently large compared to the dot-lead tunneling strength that generally only one, or occasionally two, levels are involved in transport.

When just one dot level is relevant the standard model is the Anderson impurity model (AIM).\cite{Anders1961,Ng1988,Glazma1988} Here the dot itself is described by
\begin{equation}
\label{eqn:AIM}
\Hdot^{\mathrm{AIM}}=\sum_{\sigma} \left(\epsilon_{1\sigma}+\tfrac{1}{2}U\nmso\right)\nso
\end{equation} 
where $\hat{n}_{1\sigma}=\dcreo\ddeso$ counts the $\sigma$ spin electrons on the dot level, $U$ is the 
on-level Coulomb replusion/charging energy, and $\epsilon_{1\sigma}=\epsilon_1-\sigma h$ the level energy. The latter includes a Zeeman coupling to an external magnetic field $B$ with $h=\half g\mub B$ and $\sigma = +/-$ for $\uparrow/\downarrow$-spin electrons. In the case of two relevant dot levels, the dot Hamiltonian is naturally more complex. We choose to work with the following two-level model (2LM)
\begin{equation}
\label{eqn:2levdot}
\Hdot^{\mathrm{2LM}}=\sum_{i,\sigma} \left(\epsilon_{i\sigma}+\tfrac{1}{2}U\nims\right)\nis+\up\none\ntwo-J_H \shat{1}\cdot\shat{2},
\end{equation}
which has previously been shown to capture the key physics of two-level quantum dots in the absence of a magnetic field.\cite{Logan2009} Here $\nic=\sum_\sigma\nis$ is the total number operator for level $i$ (=1,2), and $\shat{i}$ is the local spin operator with components $\hat{s}_i^{\alpha}=\sum_{\sigma,\sigma'}\dicre\sigma_{\sigma\sigma'}^{\alpha}d_{i\sigma'}^{\pd}$  ($\boldsymbol\sigma_{\sigma\sigma'}$ are the Pauli spin-$\half$ matrices). In addition to the on-level Coulomb repulsion $U$ 
(taken to be identical for levels 1 and 2 for simplicity), the model includes an interlevel Coulomb repulsion 
$U'$ plus a ferromagnetic (Hund's rule) exchange coupling of the spins of the two levels, parameterized by $J_H$.

In each case, the dot Hamiltonian is supplemented by coupling to two equivalent, noninteracting `left' and `right' leads, themselves described by $\Hl=\sum_\nu\sum_{\K,\sigma}\ek\acre\ades$ ($\nu=L,R$), where the most general tunnel coupling to the leads is of  form $\Ht=\sum_\nu\sum_{i,\K,\sigma} V_{i\K\nu} (\dicre\ades+\text{H.c.})$ (the sum over level index $i$ involving just $i=1$ in the case of the AIM). The $L$ and $R$ lead chemical potentials are 
$\mu_L$ and $\mu_R$ respectively, such that for $\mu_L\ne\mu_R$ a non-zero current flows.

Analyzing the interacting models out of equilibrium is a formidable task (see e.g. ref. \onlinecite{Anders2010} for a recent discussion) and in practice we consider the equilibrium situation. This has an immediate benefit, for the AIM Hamiltonian then reduces exactly to an effective one-lead model by defining 
$\cdes=(V_{i\K L}\adesL + V_{i\K R}\adesR)/V_{i\K}$ with $V_{i\K}^2 = V_{i\K L}^2+V_{i\K R}^2$ (with $i=1$), since the corresponding orthogonal combination of lead states is entirely decoupled from the dot. The two-level dot Hamiltonian under this transformation does not generally separate so pristinely: excepting the special case of $V_{i\K L} = V_{i\K}\cos\theta$, $V_{i\K R} = V_{i\K}\sin\theta$, the dot remains coupled to two leads.\cite{Pustil2001} However, over a wide range of parameter space the second lead couples sufficiently weakly that it may in practice be neglected on energy scales of practical interest.\cite{Pustil2001} As such, for both the AIM and 2LM we work with the effective one-lead description embodied in 
\begin{equation}
\label{eqn:leads}
\Hl+\Ht = \sum_{\K,\sigma}\ek\ccre\cdes + \sum_{i=1}^N\sum_{\K,\sigma} V_{i\K} (\dicre\cdes+\text{H.c.}).
\end{equation}
with $N=1$ for the AIM and $N=2$ for the 2LM.

We consider the standard case\cite{Hewson1993} of a symmetric, flat-band lead of half-width $D$ and density of states per orbital $\rho_0=1/(2D)$, and take $V_{i\K}\equiv V$. 
The dot-lead coupling is then embodied in the hybridization strength $\Gamma=\pi\rho |V|^2$
(with $\rho ={\cal{N}}\rho_{0}$ the total density of states, and ${\cal{N}}\rightarrow \infty$ the number of lead orbitals). When presenting results, we use dimensionless parameters defined in terms of $\Gamma$, viz.
\begin{gather}
\label{eq:rescaledparams}
  \tilde{\epsilon_{i}}=\frac{\epsilon_i}{\Gamma},\qquad
  \util=\frac{U}{\Gamma},\qquad
  \uptil=\frac{\up}{\Gamma},\\
  \jtil=\frac{J_H}{\Gamma},\qquad\notag
  \htil=\frac{h}{\Gamma}.
\end{gather}
The bandwidth $D$ is naturally taken to be the largest energy scale in the problem, and for our NRG calculations in practice we take $D/\Gamma = 100$.

To study the models described above, the central quantities of interest are the dot Green functions $G_{ij;\sigma}(\omega;h)$ [$\leftrightarrow
G_{ij;\sigma}(t;h)=-i\theta(t)\langle\{\dides(t),d_{j\sigma}^\dagger(0)\}\rangle$] with associated spectral density $D_{ij;\sigma}(\omega;h)=-\tfrac{1}{\pi}\mathrm{Im}G_{ij;\sigma}(\omega;h)$ 
($\theta(t)$ denotes the unit step function).
In the absence of an applied magnetic field, $G_{ij;\uparrow}(\omega;0) = G_{ij;\downarrow}(\omega;0)$, while for any finite $h$ the $\uparrow$- and $\downarrow$-spin Green functions are naturally distinct. In addition to these spin-resolved quantities, we will later make use of their spin-summed analogs, in particular the spin-summed spectrum 
\begin{equation}
\label{eqn:Gij}
 D_{ij}(\omega;h)=\half\sum_\sigma D_{ij;\sigma}(\omega;h).
\end{equation}
Moreover, in the case of the 2LM some of the physics is better described in terms of the symmetrized combinations of dot orbitals\cite{Logan2009}
\begin{equation}
\label{eqn:combos}
 d_{e\sigma}=\frac{1}{\sqrt{2}}(d_{1\sigma}+d_{2\sigma}), ~~~ d_{o\sigma}=\frac{1}{\sqrt{2}}(d_{1\sigma}-d_{2\sigma})
\end{equation}
from which follow the `even-even' and `odd-odd' Green functions:
\begin{align}\label{eqn:Geeoo}
  G_{ee;\sigma}(\omega)&=\half\left[G_{11;\sigma}(\omega)+G_{22;\sigma}(\omega)+ 2G_{12;\sigma}(\omega)\right]\\
  G_{oo;\sigma}(\omega)&=\half\left[G_{11;\sigma}(\omega)+G_{22;\sigma}(\omega)- 2G_{12;\sigma}(\omega)\right].
\end{align}

The connection between theory and experiment is made via the zero-bias differential conductance,
$G_c^0(T)$. For the models considered above, this is obtained exactly from the Meir/Wingreen approach\cite{Meir1992} which gives
\begin{equation}
\label{eqn:MWGc}
 G_c^0(T;h)=\frac{2e^2}{h}G_0 \int_{-\infty}^{+\infty} d\omega \frac{-\partial f(\omega)}{\partial \omega}~ N\pi\Gamma D_{ss}(\omega;h).
\end{equation} 
Here $f(\omega)=(e^{\omega/T}+1)^{-1}$ $(k_B\equiv 1)$ is the Fermi function, and $D_{ss}(\omega;h)=\tfrac{1}{N}\sum_{i,j} D_{ij}(\omega;h)$ is the spectral density of the fully symmetric impurity channel [i.e. $D_{11}(\omega;h)$ for the AIM and $D_{ee}(\omega;h)$ for the 2LM]. The dimensionless prefactor $G_0=\sin^22\theta$ reflects the relative coupling asymmetry to the right and left leads and is maximal, $G_0=1$, for equal couplings.\cite{Logan2009}

As alluded to above, present methods cannot give exact results for the non-equilibrium situation of a finite source-drain bias between the leads. While recent progress has been made in addressing this (see e.g. refs.
\onlinecite{Anders2008,Schmitt2011,Anders2010}), the methods used are much more computationally intensive and thus we make the standard approximation of neglecting the $V_{sd}$ dependence of the impurity self-energy. The result is that
\begin{equation}\label{eqn:GcVsd}
\begin{split}
&\frac{G_c(T,V_{sd};h)}{(2e^2/h)G_0}=\\
&-\int_{-\infty}^{+\infty} d\omega \left[\lambda\frac{\partial f_L(\omega)}{\partial \omega}+(1-\lambda)\frac{\partial f_R(\omega)}{\partial \omega}\right]N\pi\Gamma D_{ss}(\omega;h)
\end{split}
\end{equation} 
where $f_\nu(\omega)=f(\omega-\mu_\nu)$ with $\mu_L = \lambda eV_\mathrm{sd}$ and $\mu_R=-(1-\lambda) eV_\mathrm{sd}$. The quantity $\lambda \in (0,1)$ controls the partitioning of the voltage split $eV_\mathrm{sd}$ between the two leads, with $\lambda = \half$ corresponding to a symmetric voltage drop. In \sref{sec:experiment} we use \eref{eqn:GcVsd} to interpret a number of recent experimental results, where in particular we discuss critically the agreement between this quasi-equilibrium approximation and experiment.

Finally, our numerics are obtained from the full density matrix (FDM)\cite{Weichs2007,Peters2006} formulation of the 
NRG,\cite{Wilson1975,Krishn1980,Bulla2008} using the Oliveira discretization scheme\cite{Olivei1994} and a generalization of the self-energy method of Bulla \emph{et al.}.\cite{Bulla1998} We find it sufficient to keep $\sim 4000$ states per NRG iteration, and employ an NRG discretization parameter $\Lambda =3$.

\section{Zero-field physics and low-energy effective models}
\label{sec:zerofield}
To put our finite-field results in context, we consider briefly the zero-field physics of the two models; 
starting with the AIM, which at zero field is well understood by a range of complementary techniques (see e.g. ref.~\onlinecite{Hewson1993}).

The AIM exhibits local Fermi liquid behavior for all $\Gamma>0$, as reflected in the RG description by a single stable fixed point (FP): the strong-coupling (SC) fixed point.\cite{Wilson1975,Krishn1980} For fixed $U/\Gamma$, the dot occupancy $n_1=\langle \hat n_1\rangle$ increases continuously with decreasing $\epsilon_1$, starting close to $n_1\simeq 0$ when $\epsilon_1/\Gamma \gg 1$, and tending to a maximum of 2 when $(\epsilon_1+U)/\Gamma \ll 1$. At the point $\epsilon_1=-U/2$ the model is invariant under a particle-hole (p-h) transformation\cite{Hewson1993} and hence $n_1=1$ precisely.

When charge fluctuations are suppressed by a large $U/\Gamma\gg 1$, the dot occupancy tends toward integer values, increasing more-or-less stepwise as $\epsilon_1$ is decreased (under a gate voltage in practice, $\epsilon_{1} \propto V_{\mathrm{gate}}$). In the singly-occupied regime ($n_1\simeq 1$), the AIM reduces under a Schrieffer-Wolff transformation\cite{Schrie1966} of \erefs{eqn:AIM}{eqn:leads} to a low-energy effective Kondo model: defining a p-h asymmetry parameter $\eta = (1+2\epsilon_1/U)$, for a fixed $-1<\eta<1$ and 
$\tilde U = U/\Gamma\gg 1$ this yields
\begin{equation}
\label{eqn:kondoham}
\hat{H}_K=\sum_{\mathbf{k},\sigma}\epsilon_\mathbf{k}^{\phantom\dagger}\ccre\cdes+J\hat{\mathbf{s}}\cdot \hat{\mathbf{S}}(0)+K\sum_{\sigma}f^{\dagger}_{0\sigma} f^{\phantom\dagger}_{0\sigma}
\end{equation}
where $\hat{\mathbf{s}}$ is a spin-$\half$ operator describing the dot spin,
$\hat{S}^\alpha(0)=\sum_{\sigma,\sigma'}f_{0\sigma}^\dagger\sigma_{\sigma\sigma'}^{\alpha}f_{0\sigma'}^{\pd}$
is the conduction band/lead spin density at the dot, and
$f_{0\sigma} =\tfrac{1}{{\cal{N}}}\sum_{\mathbf{k}}c_{\mathbf{k}\sigma}$. The Kondo exchange coupling $J$ and potential scattering strength $K$ are given in terms of the original model parameters by 
\begin{equation}
\label{eqn:AIMJK}
    \rho_0 J=\frac{8}{\pi \tilde U}\frac{1}{1-\eta^2} \qquad
     \rho_0 K=\frac{2}{\pi\tilde U}\frac{\eta}{1-\eta^2},
\end{equation}
such that at p-h symmetry ($\eta =0$) the potential scattering $K =0$. Away from p-h symmetry potential scattering is non-vanishing but, from \eref{eqn:AIMJK} $K/J=\eta/4$ and hence for fixed $\eta$ the model is characterized by a single dimensionless parameter $\rho_0 J$. This in turn means that for a given $\eta$, all AIMs in the strongly interacting $\tilde U\gg1$ regime map onto the same low-energy effective Hamiltonian; and thus exhibit universal scaling of their physical properties in terms of the low-energy Kondo scale $\tk\sim D\exp[-1/(\rho_0 J)]$.\cite{Hewson1993}

The physics of the 2LM is naturally more complicated. We refer the reader to ref. \onlinecite{Logan2009} for detailed discussion, and merely summarize the key points here. When considering the model as a function of the level energies $\epsilon_1$ and $\epsilon_2$, it is more convenient to work with 
\begin{equation}
\label{eqn:xy}
  x=\epsilon_1+\half U+U', \qquad
  y=\epsilon_2+\half U+U', 
\end{equation}
since it can be shown that the model is p-h symmetric when $x=0=y$, and that its phase diagram is symmetric under reflection in the lines $y=\pm x$.\cite{Logan2009} That the phase diagram itself is non-trivial reflects the occurrence now of two stable FPs: the SC FP again, and the underscreened spin-1 (USC) FP of Nozi\'eres and Blandin.\cite{Nozier1980} As for the AIM, the SC phase is a local Fermi liquid, while the USC phase is a singular Fermi liquid\cite{mehta,Logan2009} characterized by a free spin-$\half$ on the dot with a $\ln 2$ residual entropy. 

Close to p-h symmetry ($x=0=y$) the dot levels are each singly occupied, and in the absence of coupling to the lead naturally form a spin-triplet. On coupling to the lead, this spin-1 is reduced to an effective spin-$\half$ by the underscreened Kondo effect,\cite{Nozier1980} whence a finite region surrounding the p-h symmetric point belongs to the USC phase. On moving further away from p-h symmetry [in any direction in the $(x,y)$ plane], the model eventually undergoes a quantum phase transition to the SC phase (see Fig.~5 of ref.~\onlinecite{Logan2009}). 
The transition is of Kosterlitz-Thouless (KT) type,\cite{Logan2009} the Kondo scale in the SC phase vanishing exponentially as the boundary to the USC phase is approached. This holds generically except at points of special symmetry (specifically along the line $y=x$, where the transition becomes  first-order\cite{Logan2009}).

The effective low-energy model `deep' in the underscreened triplet regime can be obtained by Schrieffer-Wolff 
on the 2LM, \erefs{eqn:2levdot}{eqn:leads}, valid formally for $U/\Gamma\gg 1$ with fixed $J_H/U$ and $U'/U$. The resulting model is spin-1 Kondo with potential scattering;\cite{Logan2009} of the same form as \eref{eqn:kondoham} but with $\hat{\mathbf{s}}$ now a spin-$1$ operator and $J, K$ given by
\begin{subequations}
\label{eqn:2LevJK}
  \begin{align}
    \rho_0 J(x,y)&=\frac{4}{\pi(\tilde U+\half \tilde J)}\left[\frac{1}{1-\eta(x)^2}+\frac{1}{1-\eta(y)^2}\right]\label{eqn:2LevJ}\\
    \rho_0 K(x,y)&=\frac{2}{\pi(\tilde U+\half \tilde J)}\left[\frac{\eta(x)}{1-\eta(x)^2}+\frac{\eta(y)}{1-\eta(y)^2}\right]
    \label{eqn:2LevK}
  \end{align}
\end{subequations}
where (with $z=x$ or $y$) the asymmetry is 
\begin{equation}
\label{eqn:asymm}
  \eta(z)=\frac{2z}{U+\half J_H}.
\end{equation}
The characteristic Kondo scale deep in the USC phase is\cite{Andrei1983} $\tk^{S=1}\sim D \exp[-1/(\rho_0 J)]$ (i.e. has the same exponential dependence on $\rho_0 J$ as for the spin-$\half$ case).

In direct analogy to the AIM, the ratio of $K$ to $J$ is a function solely of the asymmetries, conveniently expressed in terms of a quantity $\eta(x,y)$:
\begin{equation}
\label{eqn:etaxy}
 \eta(x,y) =\frac{2K(x,y)}{J(x,y)} ~~~
  = ~\frac{\eta(x)+\eta(y)}{2-\frac{[\eta(x)-\eta(y)]^2}{1-\eta(x)\eta(y)}}
\end{equation}
Sufficiently deep in the USC phase, one thus expects physical properties of the 2LM to be universal in $T/\tk^{S=1}$ 
for fixed $\eta(x,y)$; as considered further in \sref{sec:statdynam}.

\section{Field Dependent Statics and Dynamics}
\label{sec:statdynam}
We now turn to our main focus: the effect of a applied magnetic field on the AIM and 2LM. 
While much is already known for the AIM, certain aspects of its dynamics in a magnetic field\cite{Logan2001,Costi2000,Hofste2000,Hewson2006,Moore2000,Bauer2007,Bauer2007a,Zitko2009} have not
been fully understood, and in \sref{ssec:dynam} we present NRG results to clarify the situation. The 2LM model has been less widely studied, and we consider it in somewhat more detail.

\subsection{Magnetization}\label{ssec:magnetization}
It is first instructive to consider the magnetization for level $i$, here defined by
\begin{align}
m_i(h) &=\langle \hat n_{i\uparrow}\rangle - \langle \hat n_{i\downarrow}\rangle\label{eqn:mh1}\\
&=\int_{-\infty}^0 d\omega\left[ D_{ii;\uparrow}(\omega) - D_{ii;\downarrow}(\omega)\right].\label{eqn:mh2}
\end{align}
This can be determined accurately using the FDM-NRG,\cite{Weichs2007,Peters2006,Bulla2008} the complete Fock space approach circumventing known problems arising in the original NRG.\cite{Costi2000,Hofste2000}

\begin{figure}
\includegraphics{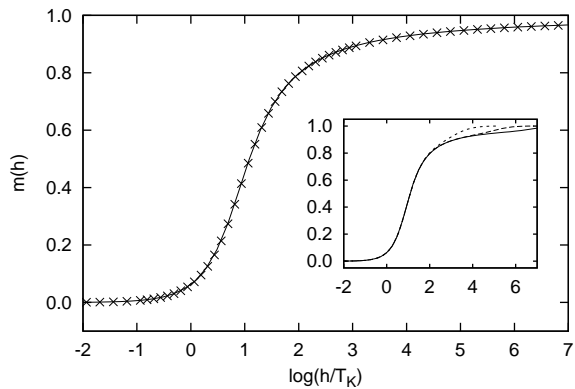}
\caption{\label{fig:BAcompare}
Magnetization of the AIM \emph{vs} $h/\tk$. \emph{Main}: $\util=-2\etil=50$. Comparison between FDM-NRG results (crosses) calculated via
\eref{eqn:mh1}, and the Bethe Ansatz result\cite{Andrei1983} (line) for the Kondo Model. \emph{Inset}: $\util=-2\etil=30$ (solid), $20$ (long dash) and $10$ (short dash) corresponding to $\tk/\Gamma\sim\expnum{1.8}{-6},\expnum{8.0}{-5}$ and $\expnum{3.1}{-3}$. Deviations from universality occur for
$h\gtrsim {\cal{O}}(\Gamma)$.
}
\end{figure}

\Fref{fig:BAcompare} shows the total dot magnetization $m(h)=m_1(h)$ \emph{vs} $\log(h/\tk)$ for the 
symmetric AIM with $\util=50$ and $\etil=-\half\util$.\cite{note:tk} The accuracy of the FDM-NRG is confirmed by the clear agreement with the exact result known from the Bethe \emph{ansatz}\cite{Andrei1983} for the Kondo model. At a field $h\sim \tk$ the magnetization rises rapidly from its zero field (Kondo-screened) value $m(0)=0$, before turning over to a slow asymptotic approach to saturation of form $m(h)\sim 1-[2\ln{(h/\tk)}]^{-1}$.
The inset to \fref{fig:BAcompare} gives results for $\util=30,20$ and $10$, showing the inevitable deviation from the universal Kondo magnetization curve at sufficiently high fields $h\gtrsim {\cal{O}}(\Gamma)$.

We now compare this behavior to that of the two-level model of \eref{eqn:2levdot}. The basic physics now reflects the destruction of the quantum phase transition occurring for $h=0$, and its replacement by a smooth crossover. In terms of FPs, the spin symmetry breaking associated with the magnetic field renders the USC fixed point unstable for all $h\ne 0$, and so ultimately \emph{all} NRG flows tend toward a SC fixed point (now supplemented by spin-dependent potential scattering).

\begin{figure}
\includegraphics{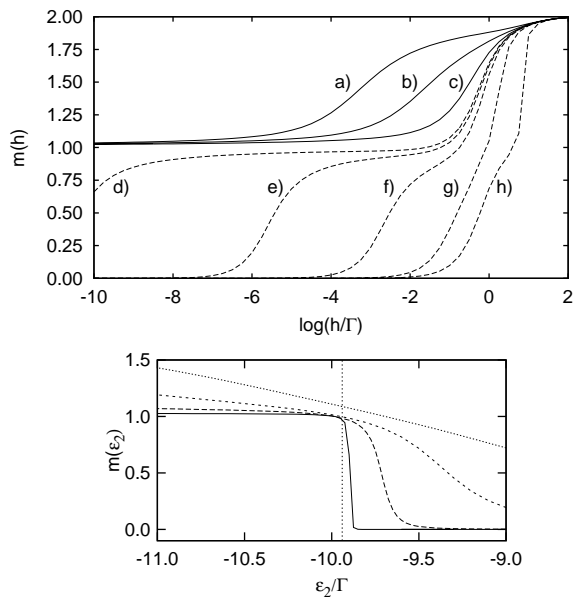}
\caption{\label{fig:magtrans}
\emph{Upper}: Total magnetization $m(h)=m_1(h)+m_2(h)$ of the 2LM with $\util=2\uptil=4\jtil=20$, $\tilde\epsilon_1=-\half \util-\uptil=-20$ and $\tilde\epsilon_2=$ (a) $-20$, (b) $-12$, (c) $-10.5$, (d) $-9.9$, (e) $-9.8$, (f) $-9.5$, (g) $-8$ and (h) $-2$. At zero field (a)--(c) correspond to the USC phase; here an infinitesimal field polarizes the underscreened impurity moment. For larger $\epsilon_2$ the system is in the SC phase at zero field, and $m(h\rightarrow 0) \rightarrow 0$.
\emph{Lower}: Magnetization near the crossover for 
$\tilde{h}= 10^{-10}$ (solid), $10^{-4}$ (long dashed), $10^{-2}$ (short dashed) and $10^{-1}$ (dotted).
The $h=0$ transition is marked by a dotted vertical line at $\etil_{2c}=-9.94..$. For $h\rightarrow 0$,  
$m$ jumps discontinuously at $\etil_{2c}$ while for finite fields this step is smeared.
}
\end{figure}

\Fref{fig:magtrans} \emph{upper} shows the total dot magnetization $m(h)=m_1(h)+m_2(h)$ for $\util=2\uptil=4\jtil=20$ and with fixed $\epsilon_1=-\half U-U'$ (i.e. $x=0$, see \eref{eqn:xy}), upon increasing
$\epsilon_2$ (or $y$) from its p-h symmetric value. In the following it is useful to bear in mind that on increasing $\epsilon_2$ at \emph{zero field}, the model undergoes the quantum phase transition from USC to SC at a critical $\tilde\epsilon_{2\mathrm{c}}\simeq -9.94$.

Curves (a) to (c) in \fref{fig:magtrans} correspond to $\epsilon_2 < \epsilon_{2\mathrm{c}}$ and hence the USC phase at $h=0$. At finite field these curves show $m(h)\rightarrow 1$ as $h\rightarrow 0$: an infinitesimal field fully polarizes the free spin-$\half$ local moment associated with the USC fixed point (for $h=0$ identically, by contrast, the magnetization vanishes by symmetry). On increasing $h$, the magnetization increases monotonically, crossing over towards $m(h)=2$ on the scale $h\sim\Tkone$ as the field destroys the underscreened Kondo effect and singles out the $S_z=+1$ component of the dot triplet state.\cite{note:tkspin1}
As shown in ref.~\onlinecite{Logan2009}, $\Tkone$ increases upon moving away from the center of the USC phase, hence the higher field required to destroy the underscreened Kondo effect for (c) compared to (a).

For larger level separations, $\epsilon_2 > \epsilon_{2\mathrm{c}}$ (curves (d) to (h)), the zero-field phase is SC. Here the low-field behavior more closely resembles that of \fref{fig:BAcompare}. At zero-field the dot is fully screened by the lead, and remains essentially so until $h$ on the order of the SC phase Kondo scale, $\tk$; above which the spin-$\half$ Kondo effect is progressively destroyed, and $m(h)$ crosses over to $m(h) \sim 1$ associated with a spin-polarized spin-$\half$ on the dot. As in curves (a)--(c), increasing the field further then causes a second marked increase in $m(h)$ when the $S_z=+1$ component of the two-electron triplet state is favored.

\Fref{fig:magtrans} \emph{lower} shows the magnetization at various fixed values of $h/\Gamma$ as a function of $\tilde\epsilon_2$ (focussing on the vicinity of the zero-field transition at $\etil_{2c}$). At any finite field, the magnetization decreases monotonically with increasing $\tilde\epsilon_2$, and as $h/\Gamma\to 0$ the curve approaches the step function $m(\epsilon_2)\to\theta(\etil_{2c}-\etil_{2})$.\cite{Pustil2006} In the absence of the field, however, the magnetization naturally vanishes, and hence the limit of $h\rightarrow 0^+$ and $h=0$ are quite distinct.  

\begin{figure}
\includegraphics{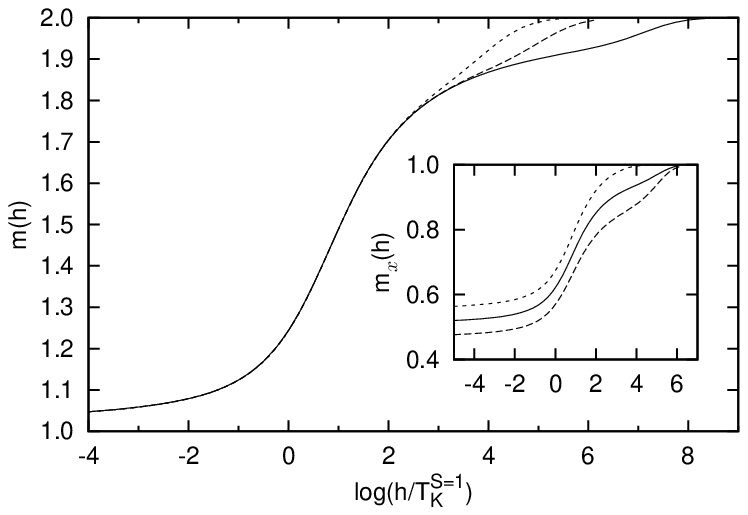}
\caption{\label{fig:Mag}
Magnetization $m(h/\Tkone)$ for the two-level model at p-h symmetry ($\epsilon_1=\epsilon_2=-\tfrac{U}{2}-U'$),
deep in the USC phase. $\util=2\uptil=4\jtil=30$ (solid), 20 (long dash) and 15 (short dash), corresponding to $\Tkone/\Gamma=\expnum{1.0}{-6},\expnum{7.7}{-5}$ and $\expnum{6.8}{-4}$.
As with the AIM, $m(h/\Tkone)$ exhibits universal scaling.\cite{Andrei1983,Furuya1982}
\emph{Inset}: Comparison of $m_x(h)$, $x\in\{1,e,o\}$ (solid, long dash, short dash), for the $\util=20$ case. 
}
\end{figure}

As for the spin-$\half$ Kondo effect in \fref{fig:BAcompare}, the magnetization deep in the USC phase (where the low-energy effective model is spin-1 Kondo) is a universal function of $h/\Tkone$. \Fref{fig:Mag} illustrates 
scaling of the magnetization for $\util=2\uptil=4\jtil=30$, $20$ and $15$ at the p-h symmetric point $(x,y)=(0,0)$ where $\expn{i}=1$ for $i\in\{1,2,e,o\}$ (see \eref{eqn:combos}). The main figure shows $m(h)$, the total impurity magnetization. Results for different values of the bare parameters clearly display scaling, onto a different universal form than for the AIM.\cite{Andrei1983,Furuya1982}

\Fref{fig:Mag} (\emph{inset}) shows also the magnetization of the even and odd impurity orbitals 
(such that $m(h)=m_{e}(h)+m_{o}(h)$), for the $\tilde U=20$ case. The $o$-orbital is clearly
polarized to a greater extent than the $e$-orbital by an infinitesimal field, and $m_o(h)$ reaches saturation more quickly than $m_e(h)$ (or indeed $m_1(h)$). This reflects the fact that the $o$-orbital does not couple directly to the conduction band,\cite{Logan2009} only interacting with it via the $e$-orbital which couples directly; the 
$o$-orbital as such contributing more to the local moment
than the $e$-orbital. 

\begin{figure}
\includegraphics{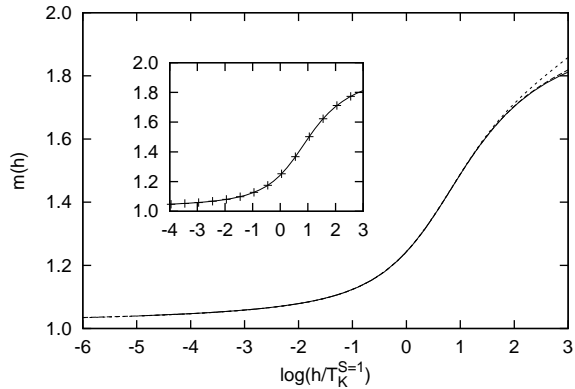}
\caption{\label{fig:awayMag}
Magnetization away from p-h symmetry, deep in the USC phase. \emph{Main}: $m(h)=m_1(h)+m_2(h)$ for the same systems as \fref{fig:Mag} but with $\eta(x)=0$ and $\eta(y)=0.5$. Clear universality is observed. \emph{Inset}: The $\util=30$ case on the $y=-x$ line, for $x=0$ (line) and $x/\Gamma=-6$ (crosses).\\
The resultant universal $m(h)$ is found to be the same in \emph{both} cases, i.e.\ to be independent of asymmetry.
}
\end{figure}

The situation deep in the USC phase, but away from p-h symmetry, is illustrated in \fref{fig:awayMag}. As mentioned in \sref{sec:zerofield}, universal behavior of $m(h)$ is expected for systems with different bare parameters, at least for fixed asymmetry $\eta(x,y)$ (i.e.\ from \eref{eqn:asymm} the same ratio of potential scattering $K$ to Kondo coupling $J$). To this end consider first the line $y=-x$, for \emph{all} points on which $\eta(x,y) =0$ (\erefs{eqn:etaxy}{eqn:asymm}). \Fref{fig:awayMag} (inset) shows the universal $m(h)$ at the p-h symmetric point $x=0=-y$ (line) considered also in \fref{fig:Mag}, compared to that obtained some distance away from p-h symmetry at $x=6\Gamma=-y$ (crosses). The two scaling curves clearly coincide.

The main panel of \fref{fig:awayMag} illustrates universality for non-vanishing asymmetry, showing $m(h)$ \emph{vs} $h/\Tkone$ for the same $\util=2\uptil=4\jtil$ values as \fref{fig:Mag}, but now with $\eta(x)=0$ and $\eta(y)=0.5$ (i.e.\ $\eta(x,y)\simeq 0.29$). The three curves scale perfectly in the universal regime, beginning to deviate only at high fields $h\sim \mathcal{O}(\Gamma)$. Moreover, the resultant universal $m(h)$ is found numerically to be identical to that arising for $\eta(x,y)=0$, and as such thus appears to be \emph{independent} of asymmetry $\eta(x,y)$; a result we have further confirmed for a wide range of $\eta$-values.

\subsection{Field Dependent Dynamics}\label{ssec:dynam}
We turn now to the field dependence of single-particle dynamics for the AIM and 2LM. Much is already known \cite{Logan2001,Costi2000,Hofste2000,Hewson2006,Moore2000,Bauer2007,Bauer2007a,Zitko2009,Zitko2011} about the former case, but it serves as a useful comparison to the 2LM and both are experimentally relevant. The spin-resolved impurity spectrum is first considered, with a two-fold focus: the field-induced redistribution of weight in the Hubbard satellites, and the shift of the spectral maximum from zero.

\begin{figure}
\includegraphics{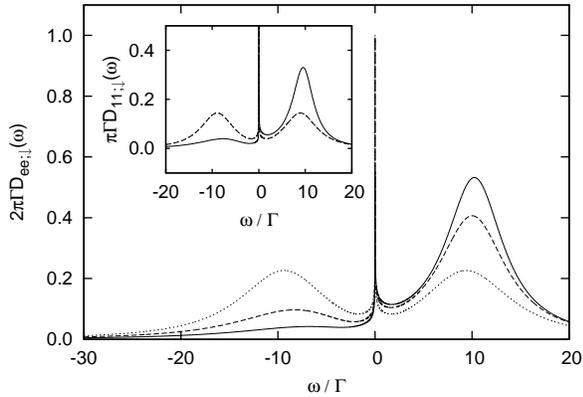}
\caption{\label{fig:Hubbard}
Redistribution of spectral weight in the Hubbard bands. \emph{Main}: 2LM with $\util=2\uptil=4\jtil=20$ at p-h symmetry, with $h/\omega_K^{S=1} =0$ (dotted), $\expnum{1}{-6}$ (dashed) and $1$ (solid), $\omega_K^{S=1}$ being the Kondo scale defined as the HWHM of the zero-field Kondo resonance. \emph{Inset}: AIM with $\util=20$ also at p-h symmetry with $h/\omega_K =0$ (dotted line), $\expnum{1}{-6}$ (dashed) and $1$ (solid). The result for $h/\omega_K =\expnum{1}{-6}$ is coincident with that for zero field on the scale shown.
}
\end{figure}

\Fref{fig:Hubbard} shows results for $D_{ss;\downarrow}(\omega;h)$ and a range of fields $h\geq 0$, for both the AIM (inset) and the 2LM. The level energies and interaction strengths have been chosen so that both models are 
deep in the Kondo regime (for the AIM) or underscreened triplet (2LM), and are p-h symmetric
(such that $D_{ss;\uparrow}(\omega;h)=D_{ss;\downarrow}(-\omega;h)$). In each case
the familiar three peak structure is evident: upper and lower Hubbard satellites due to local charge excitations on the impurity, and a central low-energy Kondo resonance. We denote the half-width at half-maximum of the $h=0$ Kondo resonance by $\omega_K$: the low-energy Kondo scale, proportional to the Kondo temperature $\tk$.

 In both cases, increasing the applied field causes spectral weight to be redistributed from the lower to the upper Hubbard satellite, corresponding to the destabilisation of $\downarrow$-spin electrons on the dot. The striking difference between the two is that for the 2LM (main figure), a significant redistribution occurs upon introducing an infinitesimal field (e.g. $h/\omega_K^{S=1}=\expnum{1}{-6}$), whereas for the AIM this occurs only when $h/\omega_K\sim {\cal{O}}(1)$ (inset). This reflects directly the behavior of the magnetization in \fref{fig:magtrans} (see \eref{eqn:mh2}): the free spin associated with the USC FP is fully polarized by an infinitesimal field, while a finite $h\sim \omega_K$ is required to disrupt the Kondo singlet associated with the SC FP of the AIM.

The above high-frequency behavior is relatively straightforward compared to that at lower energies
$\omega\sim \omega_K$; as now addressed, beginning with the AIM. The low-frequency behavior of the AIM spectrum in a magnetic field has received significant attention using various techniques,\cite{Costi2000,Hofste2000,Moore2000,Logan2001,Hewson2006,Bauer2007,Bauer2007a,Zitko2009,Zitko2011} yet there is still some disagreement in the literature. Here we present results from accurate NRG calculations, with the aim of clarifying the issue. 

\begin{figure}
\includegraphics{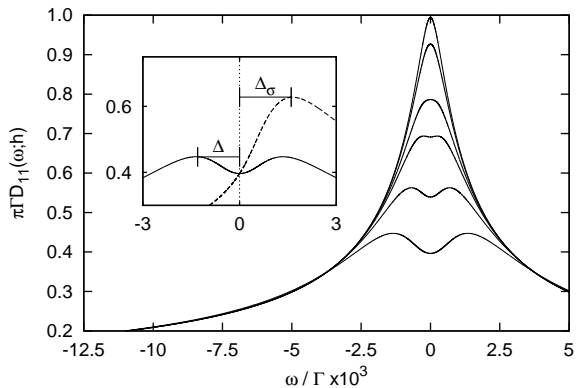}
\caption{\label{fig:Peaks}
Kondo resonance splitting in the spin-summed $D_{11}(\omega;h)$ for the AIM, on application of a magnetic field. $\util=20$, $\etil=-10$ and (top to bottom) $h/h_C=0$, $1/2$, $1$, $4/3$, $2$ and $3$ with $h_C=0.27\omega_K$ and $\omega_K/\Gamma=\expnum{1.7}{-3}$. \emph{Inset}: Kondo peaks in $D_{11}(\omega;h)$ (solid) and $D_{11;\downarrow}(\omega;h)$ (dashed).
}
\end{figure}

At zero field the Kondo resonance at p-h symmetry is centered on $\omega=0$, symmetric to reflection about $\omega=0$, and satisfies the Fermi liquid pinning condition $\pi\Gamma D_{11;\sigma}(\omega=0)=1$. Introduction of a finite $h$ is well known to shift the resonance in $D_{11;\sigma}(\omega;h)$
away from $\omega=0$ and diminish its height.\cite{Costi2000} We define $\Delta_\sigma$ as the magnitude of this shift, as shown in the inset of \fref{fig:Peaks}.

The spin-summed spectrum $D_{11}(\omega;h)$ ($=\tfrac{1}{2}\sum_{\sigma} D_{11;\sigma}(\omega;h)$) is distinct from the individual $D_{11;\sigma}(\omega;h)$, since  the $\sigma=\uparrow$ and $\downarrow$ Kondo resonances are 
shifted in opposite directions by the field. At sufficiently high fields, the two resonances are far apart and 
$D_{11}(\omega)$ contains two peaks separated by $2\Delta$ (see \fref{fig:Peaks}, inset). As $h$ is reduced, these 
peaks approach each other and are known\cite{Costi2000} to coalesce at a field we denote $h_C$ (\fref{fig:Peaks}, main). Our FDM-NRG calculations yield a universal value $h_C\simeq 0.27\omega_K$ in the Kondo regime ($U/\Gamma\gg 1$). In terms of the quasiparticle weight $Z=[1-\partial\Sigma_{11}(\omega=0)/\partial\omega]^{-1}$, easily extracted from FDM-NRG results for the dot self-energy $\Sigma_{11}(\omega)$, we obtain $h_C\simeq0.40Z\Gamma$. This is in good agreement with the exact result of ref.~\onlinecite{Hewson2005}, $h_C/Z\Gamma=0.39\ldots$.

We have performed accurate NRG calculations to determine the universal scaling behavior of $\Delta_\sigma/h$ and $\Delta/h$ as a function of $h/\omega_\mathrm{K}$. Before discussing these results it is worth explaining the calculational procedure itself. We find that to calculate $\Delta_\sigma$ and $\Delta$ accurately over a wide range of $h/\omega_K$, it is necessary to combine results from different values of $\util$. For a given $\util$, one cannot obtain universal results for arbitrarily high $h/\omega_K$, because universality arises only when $h$ is much smaller than the non-universal scales $\Gamma$ and $U$. Since $\omega_K$ is small but finite for a given $\util$, there will always be a (large) $h/\omega_K$ at which $h$ itself becomes comparable to the non-universal scales and the results then deviate from universality.

Since $\omega_K$ \emph{decreases} exponentially with increasing $\util$, this might suggest working with a very large $\util$, for then one can reach very high values of $h/\omega_K$ before $h$ itself becomes non-universal. However this is subject to a second problem, at the opposite end of the field scale. The energies that enter the Hamiltonian involve combinations of $h$, $U$ and $\Gamma$, and the double-precision arithmetic used in NRG thus places a lower limit on the size of $h$ relative to $U$ and $\Gamma$. If $\omega_K$ is too small, then low values of $h/\omega_K$ shift the dot energy levels by so little that they cannot be accurately represented in double precision. 

As such, for a given $\util$ there is a range of fields encompassing in practice around 4-5 orders of magnitude, over which the universal scaling curve can be determined by the NRG. By combining results for different values of $\util$ the full scaling curve can then be built up, and by choosing $\util$s such that the calculations overlap one can obtain a measure of the accuracy of the calculation. 

\begin{figure}
\includegraphics{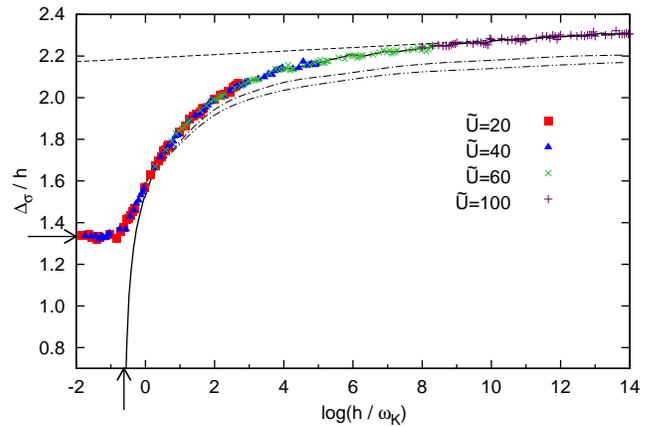}
\caption{\label{fig:AndSplit}
(Color online) Universality in the field-dependence of the spectral maximum, $\Delta_\sigma(h)$, in $D_{11;\sigma}(\omega;h)$ for the symmetric AIM. See text for discussion. Horizontal arrow indicates the exact low-field asymptote $\Delta_{\sigma}/h =\tfrac{4}{3}$. Dashed line shows the high-field form $a\log{(h/\omega_K)}+c$.
Solid line shows $\Delta(h)$, the position of the maxima in the spin-summed $D_{11}(\omega;h)$; the vertical arrow indicates the exact $h_{C}$.
}
\end{figure}

The points in \fref{fig:AndSplit} show the resultant $\Delta_{\sigma}(h)/h$ obtained from a series of NRG calculations for $\util=20$, $40$, $60$ and $100$.  Results for different values of $\util$ indeed overlap when plotted \emph{vs} $h/\omega_\mathrm{K}$, indicating universal scaling behavior. At low field $\Delta_\sigma/h\to4/3$ as $h\rightarrow 0$, recovering the exact result from Fermi liquid theory.\cite{Logan2001,Zhang2010} 
The splitting $\Delta_\sigma/h$ increases with $h/\omega_\mathrm{K}$, undergoing a rapid crossover around $h/\omega_\mathrm{K}\sim 1$ and tending asymptotically to the limiting form $\Delta_\sigma/h\sim a\log(h/\omega_K) + c$, which behavior agrees with results obtained from the local moment approach.\cite{Logan2001}

We believe the low-$h/\omega_\mathrm{K}$ behavior of the points in \fref{fig:AndSplit} to be numerically exact, having repeated our calculations significantly more accurately and obtained the same results.  The numerics also agree with recent NRG calculations\cite{Zitko2011} performed in the narrow region $-0.9\lesssim \log(h/\omega_\mathrm{K})\lesssim 0.6$.  As the field (and hence location of the spectral maximum) increases further, however, it becomes progressively more difficult to obtain accurate NRG results for $\Delta_\sigma$. This is a direct consequence of the broadening procedure employed to obtain NRG spectra: broadening is necessarily performed on a logarithmic scale due to the inherent logarithmic discretization of the technique, so sharp spectral features at finite frequencies become increasingly difficult to resolve as they move away from $\omega=0$. The problem can be resolved to some extent by using $z$-averaging\cite{Olivei1994} and calculating the self-energy directly\cite{Bulla1998}, but presently available computing power limits the extent to which this approach can be pushed. In \fref{fig:AndSplit} the points were obtained by averaging results from 10 $z$s, with a broadening parameter\cite{Bulla2008} $b=0.1$. Increasing the number of $z$s to 20 and working with $b=0.07$ and $0.05$ gives the dot-dashed and dot-dot-dashed lines in \fref{fig:AndSplit}. The results are clearly sensitive to the broadening at high fields, 
although in each case they show qualitatively similar high-field behavior. \\

\begin{figure}
\includegraphics{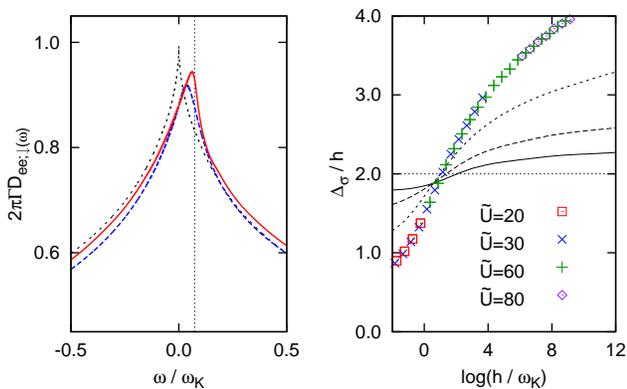}
\caption{\label{fig:TripSplit}
(Color online) Field-induced shift of the underscreened Kondo resonance in the ph-symmetric 
2LM model. \emph{Left panel}: the spectrum $D_{ee;\downarrow}(\omega)$ for $h=0$ (short dash line), and for $h/\omega_\mathrm{K}^{S=1}=0.366$ with two different broadening parameters: $b=0.7$ (dashed line) and $b=0.2$ (solid line). \emph{Right panel}: $\Delta_\sigma/h$ as a function of $h/\omega_\mathrm{K}^{S=1}$ for $b=0.7$ (points), $b=0.4$ (short-dashed), $b=0.2$ (long-dashed) and $b=0.1$ (solid), as discussed in text. The results appear to be converging to $\Delta_{\sigma}/h = 2$ for all $h/\omega_\mathrm{K}^{S=1}$; this limit is marked as a vertical dotted line in the left panel for the case $h/\omega_\mathrm{K}^{S=1}=0.366$.
}
\end{figure}

\Fref{fig:TripSplit} (right) shows analogous results for the spectral shift in $D_{ee;\sigma}(\omega;h)$ for the two-level model at p-h symmetry. Here we find the results to be even more sensitive to the NRG broadening procedure. The points show the splitting obtained from averaging 10 $z$s with $b=0.7$ (using four different bare values of $U/\Gamma$ as before), while the short-dashed, long-dashed and solid lines are from 20 $z$s with $b=0.4$, $0.2$ and $0.1$, respectively. As the accuracy of the calculation increases the splitting appears to be approaching $\Delta_\sigma/h=2$ \emph{for all} $h$, in marked contrast to the behavior of the AIM in \fref{fig:AndSplit}. 

To pursue this further, the left-hand panel of \fref{fig:TripSplit} shows $D_{ee;\downarrow}(\omega;h)$ for a representative low-field case, $h/\omega_\mathrm{K}^{S=1}\simeq 0.0366$: the long-dashed line shows the spectrum obtained with broadening $b=0.7$, while the solid line shows the $b=0.2$ result. The figure clearly illustrates the sensitivity of the finite-$h$ spectrum to the value of $b$ employed, and in line with our conjecture above it appears that in the limit $b\to 0$ the peak position would lie at $2h$ (marked as a vertical dotted line in the figure). 

The short-dashed line in \fref{fig:TripSplit} shows also the corresponding $h=0$ spectrum for comparison, the form of which (including its zero-frequency cusp) has been discussed previously.\cite{Koller2005,Logan2009} 
It is reasonable to conjecture that the finite-$h$ spectrum has a qualitatively similar form but shifted so that the cusp occurs at $\omega=2h$, although at present is is not possible to confirm or refute this using currently feasible NRG calculations.

We conclude here with a point pursued further in \sref{sec:experiment}. Our discussion has concerned purely the 
question: how does the \emph{equilibrium spectrum} evolve with magnetic field in single and two-level dots? Here we have deliberately not related the equilibrium spectrum to the finite-bias conductance, because \eref{eqn:GcVsd} is 
approximate and (as shown explicitly later in relation to recent experiments) can give quantitatively wrong 
results for field strengths in excess of a few Kondo scales.\cite{Hewson2006} The figures shown here should \emph{not} therefore be translated naively into quantitative predictions of conductance splittings. The only predictions for experiment that can currently be made with real certainty are those involving the zero-bias conductance. These are now discussed.

\subsection{Zero-bias conductance}
\label{ssec:zerobiascond}
\begin{figure}
\includegraphics{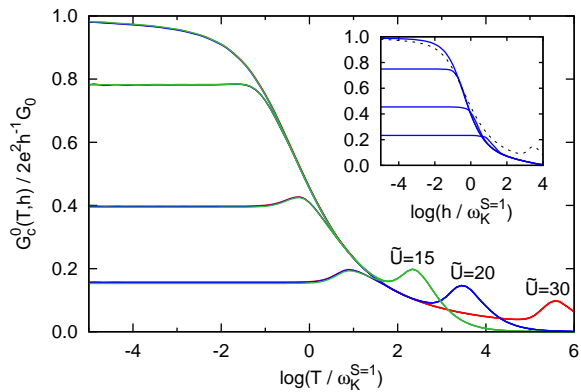}
\caption{\label{fig:CondScaling}
Scaling of zero-bias conductance for the 2LM at finite temperature and field.
 Top to bottom: $h/\omega_K^{S=1}=0,0.1,1$ and $10$. $U=2U'=4J$ with  $\util=30,20$ and $15$ (red, blue and green) corresponding to $\omega_K/\Gamma=\expnum{2.06}{-5},\expnum{1.64}{-3}$ and $\expnum{1.47}{-2}$. \emph{Inset}: $G_{c}^{0}(T,h)$ \emph{vs} $h/\omega_K^{S=1}$ for $\util=20$ and $T/\omega_K^{S=1}=0,0.1,1,10$. $G_{c}^{0}(T,h=0)$ \emph{vs} $T/\omega_{K}^{S=1}$ (dashed), as in main figure, is shown for comparison.}
\end{figure}

\Fref{fig:CondScaling} illustrates universality in the zero-bias conductance for the 2LM, as functions of $h/\omega_K^{S=1}$ and $T/\omega_K^{S=1}$. For specificity we consider the p-h symmetric point $(x,y)=(0,0)$ (which
applies also to points along the line $y=-x$ deep in the USC phase, see \sref{sec:zerofield} ). The interactions are set at $U=2U'=4J_\mathrm{H}$, and different values of $\tilde{U}=U/\Gamma$ are considered. For fixed $h/\omega_K^{S=1}$ (the values $0$, $0.1$, $1$ and $10$ are shown explicitly in \fref{fig:CondScaling}) the zero-bias conductance $G_c^0(T,h)$ is seen to be universal in $T/\omega_K^{S=1}$, as evident from clear scaling collapse of the different $\tilde{U}$ curves. Scaling naturally breaks down at non-universal scales $T\sim \min(\Gamma, U)$, and for $T \sim U$ the curves show peaks associated with incoherent sequential tunneling. 

Notice that at finite temperature for a given, sufficiently large $h/\omega_K^{S=1}$ (in excess of $\sim 0.1$ in \fref{fig:CondScaling}) there is a universal peak in the zero-bias conductance at a temperature $T\sim h$. This is analogous to the peak at finite frequency in the $T=0$ equilibrium spectrum, \sref{ssec:dynam}. Yet here the peak exists in a quantity that is both directly measurable by experiment and calculable exactly by theory. Until theory is able to capture accurately the non-equilibrium conductance as a function of source-drain bias, we suggest that the field-dependence of \emph{this} peak in the \emph{zero-bias} conductance, and more generally the $h$- and $T$-dependence of $G_{c}^{0}$, be touchstones by which the universality of experiment is assessed.

The inset to \fref{fig:CondScaling} shows another way of viewing the universal conductance curves. Here we fix $T/\omega_K^{S=1}$ (at values $0$, $0.1$, $1$ and $10$, top to bottom) and vary the magnetic field $h/\omega_K^{S=1}$ over many orders of magnitude. Notice that there is no incoherent peak at large $h$, in contrast to that in the $T$-dependence for fixed $h$. This is because at large fields the dot is completely spin polarized, and its conductance thus weak.

Before moving to particular experiments, we consider specifically the $T=0$ zero-bias conductance. For $h=0$ this is related to the scattering phase shift, $\delta$, via \eref{eqn:MWGc} and the relation 
$2\pi\Gamma D_{ee}(0)=\sin^2\delta$. In previous work\cite{Logan2009} we derived an exact Friedel-Luttinger sum rule
\begin{equation}
\label{eqn:FriedLutt}
\delta=\frac{\pi}{2}n_{\text{imp}}+I_L
\end{equation}
relating $\delta$ to the excess charge induced by the impurity, $n_{\text{imp}}$ (equivalent to $\langle\hat{n}_{1}+\hat{n}_{2}\rangle$ in the infinite bandwidth limit), and the Luttinger integral $I_L$ defined by
\begin{equation}
\label{eqn:il}
I_L = \Im \Tr \int_{-\infty}^0 d\omega ~\frac{\partial \mathbf{\Sigma}(\omega)}{\partial \omega} \mathbf{G}(\omega).
\end{equation}
We showed\cite{Logan2009} that while $I_L = 0$ as usual for the screened Fermi liquid phase, $|I_L|=\pi/2$ is by contrast characteristic of the USC phase (\emph{regardless} of the bare model parameters), reflecting the lack of adiabatic continuity of the USC phase to the non-interacting limit. 

On applying a magnetic field, the analysis of ref. \onlinecite{Logan2009} readily generalizes to
the case of broken spin symmetry. Now one has 
$2\pi\Gamma D_{ee;\sigma}(0; h) =\mathrm{sin}^{2}\delta_{\sigma}$, with
separate phase shifts for $\sigma =\uparrow,\downarrow$ of form
\begin{equation}
\label{eqn:FriedLuttSpin}
\delta_\sigma=\pi n_{\mathrm{imp},\sigma}+I_{L\sigma}
\end{equation}
where
\begin{equation}
\label{eqn:ils}
I_{L\sigma} = \Im \Tr \int_{-\infty}^0 d\omega ~\frac{\partial \mathbf{\Sigma}_\sigma(\omega)}{\partial \omega} \mathbf{G}_\sigma(\omega),
\end{equation}
and such that\cite{Pustil2006} (via \eref{eqn:MWGc})
\begin{equation}
\label{eqn:condhpos}
\frac{G_c^0(T=0)}{(2e^2/h)G_0}=\half\left[\sin^2\delta_\uparrow+\sin^2\delta_\downarrow\right].
\end{equation}
As mentioned in \sref{ssec:magnetization} the USC fixed point is unstable for any $h>0$, which means that all NRG flows terminate at the SC fixed point. Since the SC fixed point is characteristic of adiabatic continuity to the non-interacting limit, for any finite $h$ one would expect the two Luttinger integrals $I_{L\sigma}$ to vanish.  This we have indeed confirmed by direct numerical calculation.

The Luttinger integrals thus change \emph{discontinuously} on introducing an arbitrarily small magnetic field at any point within the USC phase. One naturally then wonders whether this has consequences for the conductance. To answer this one can write the phase shifts in terms of the excess charge and magnetization\cite{Pustil2006} defined by
\begin{equation}
\begin{split}
\label{eqn:nmdef}
n &= n_{\mathrm{imp},\uparrow} + n_{\mathrm{imp},\downarrow}\\ m &= n_{\mathrm{imp},\uparrow} - n_{\mathrm{imp},\downarrow}
\end{split}
\end{equation}
(with $n\equiv \langle\hat{n}_{1}+\hat{n}_{2}\rangle$ and $m\equiv m_{1}+m_{2}$ in the infinite bandwidth limit),
from which \erefs{eqn:FriedLuttSpin}{eqn:condhpos} give 
\begin{equation}
\frac{G_c^0(T=0)}{(2e^2/h)G_0}=\half\left[1-\cos(\pi n(h))\cos(\pi m(h))\right]\label{eqn:condA}
\end{equation}
for \emph{any} point in the $(x,y)$ plane when $h>0$. But at points corresponding to the USC phase at zero field,  
$m(h=0^+) = 1$ (see e.g. \fref{fig:Mag}), i.e. it too jumps discontinuously on introducing an infinitesimal field. Substituting this into \eref{eqn:condA} gives
\begin{equation}
\frac{G_c^0(T=0)}{(2e^2/h)G_0}=\cos^2\left(\frac{\pi n}{2}\right)\;\;\;(h=0^+)\label{eqn:condB}
\end{equation}
which is precisely the conductance obtained\cite{Logan2009} in the USC phase for $h=0$. In other words, although both the Luttinger integrals and magnetization change discontinuously in the USC phase on applying a field---and hence the cases $h=0$ and $h=0^+$ are different---the conductance itself contains no signature of these abrupt changes.

\section{Experimental results}\label{sec:experiment}
\subsection{Semiconductor quantum dots}\label{ssec:liu}

Turning now to experiment, we begin by considering the work of Liu \emph{et. al.} in 
ref.~\onlinecite{Liu2009}, where the magnetic field dependence of the spin-$\half$ Kondo effect 
was measured in a GaAs device. In the experiment the gates were adjusted to produce two different realizations of a quantum dot from the same device, referred to as configurations I and II, with different dot-lead tunnel barriers.\cite{Liu2009}

We adopt the simplest theoretical model of the device: the single AIM in \eref{eqn:AIM}, and parameterize it using 
experimental data.\cite{Liu2009} At $T=0$ the equilibrium model is characterized by the two dimensionless parameters
\cite{note:bw}
$U/\Gamma$ and $\epsilon_1/\Gamma$, with the experimental $U=1.4\text{ meV}$.\cite{Liu2009} The level energy $\epsilon_1$ is as usual taken to depend linearly on the applied gate voltage $V_\mathrm{g}$: we write $\epsilon_1+U/2 = \alpha e\delta V_\mathrm{g}$, with $\delta V_\mathrm{g}$ the difference between the experimental gate voltage and its value in the center of the Coulomb blockade valley, and $\alpha$ a dimensionless constant.
Finally, comparison to experimental splittings at finite bias requires the dimensionless quantity $\lambda$ (\sref{sec:models}), that controls the partitioning of the source-drain bias $V_\mathrm{sd}$ between the leads.

The values of $U/\Gamma$ and $\alpha$ appropriate to experiment could in principle be obtained by comparing experimental and theoretical curves for $\tk / \tkn$ versus $\epsilon_1 + U/2$ over a sufficiently wide $\delta V_{G}$ range, where $\tkn$ is the Kondo scale at the center of the Coulomb valley (i.e. $\epsilon_1 + U/2 = 0$). 
We find however that the range of available data in ref.~\onlinecite{Liu2009} is insufficient to determine $U/\Gamma$
reliably in this way, since near the middle of the Coulomb valley where the experimental results have been obtained, the functional form of the theoretical Kondo scale depends only on the ratio $\alpha^{2} U/\Gamma$, and hence 
$U/\Gamma$ and $\alpha$ cannot be separately obtained. We have therefore used both the zero- and finite-field behavior to parameterize the model, choosing the best values of $U/\Gamma$, $\alpha$ and $\lambda$ to agree with the available experimental data. After analyzing a wide range of parameter space we obtain the values shown in Table~\ref{tab:results}.

\begingroup
\squeezetable
\begin{table}
\begin{ruledtabular}
\begin{tabular}{lllll}
Configuration &$U/\Gamma$ & $\alpha$ & $\lambda$ & $\tilde{T}^{0}_{K} / $K\\
\hline
I & 8.0 & 0.020 & 0.7 & 0.2\\
II & 7.2 & 0.017 & 0.65 & 0.3
\end{tabular}
\end{ruledtabular}
\caption{\label{tab:results} Parameters obtained for the two dot configurations of ref.~\onlinecite{Liu2009}, by comparison to NRG results.}
\end{table}
\endgroup

Before showing our NRG results, we comment further on the origin
of these parameters. The values of $U/\Gamma$ and $\lambda$ were determined first, simply by optimal fitting 
to the finite-field data at the center of the Coulomb blockade valley (shown in \fref{fig:ExpSplit} below), 
using the approximate \eref{eqn:GcVsd}. Then to obtain $\alpha$, we compared the experimental $\tk/\tkn$ versus gate voltage to our corresponding theoretical results, themselves taken from explicit NRG calculations.
We observe that the values of $\alpha$ so obtained are in line with the experimental estimate\cite{Liu2009} 
$\alpha \simeq 0.024$. Noting that the quantity denoted `$\Gamma$' in ref.~\onlinecite{Liu2009} is $2\Gamma$ here, the ratios of $U/\Gamma$ determined therein are $5.3$ and $4.0$ for configurations I and II, respectively. Our values are a little larger than these, contributory factors being: a) that in determining $\Gamma$ from the widths of the charging peaks one must bear in mind their many-body broadening,\cite{Logan1998, Bickers1987} 
which typically gives them a half-width at half-maximum of around $1.5-2\Gamma$ (rather than $\Gamma$, as arises in the non-interacting limit); and b) fitting Kondo scales to the Haldane formula used in ref.~\onlinecite{Liu2009} underestimates $U/\Gamma$, since it applies asymptotically in the limit $U/\Gamma\gg 1$. Given the $U/\Gamma$, and the experimental $U=1.4\text{ meV}$\cite{Liu2009}, we then calculate the Kondo temperatures 
$\tilde{T}^{0}_{K}$ as shown in Table~\ref{tab:results}, with $\tilde{T}^{0}_{K}$ defined (as in experiment\cite{Liu2009}) such that $G_c^0(\tilde{T}^{0}_{K},0)/G_0 = e^2/h$ at $\epsilon_1 = -U/2$.
Given the sensitivity of absolute Kondo scales to the bare model parameters, our values are in 
good agreement with the experimental estimates of $0.3$ K and $0.63$ K (configurations I and II respectively).\cite{Liu2009}

To add further support to these parameters we note that a consistent, independent determination of $\lambda$ can be obtained from the experimental conductance map, Fig. 1a of ref.~\onlinecite{Liu2009}. The slopes of the diagonal sequential tunneling peaks, when plotted with $V_\mathrm{sd}$ as the horizontal axis, are readily shown\cite{JCPCNT2009} to be proportional to $1/\lambda$ and $1/(1-\lambda)$, and hence their ratio
yields $\lambda/(1-\lambda)$. From the experimental data\cite{Liu2009} we extract $\lambda\approx 0.7$, in agreement with our values listed above.

\begin{figure}
\includegraphics{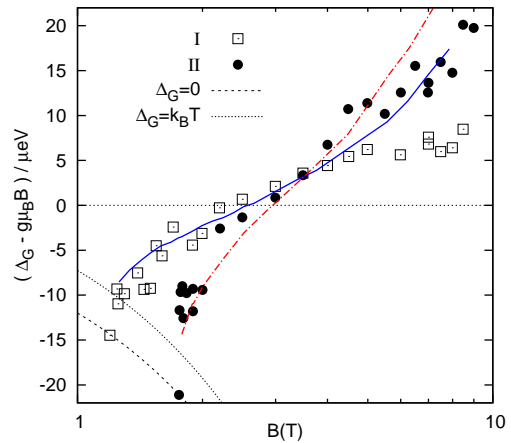}
\caption{\label{fig:ExpSplit}
(Color online) Splitting of the Kondo conductance peak on application of a magnetic field, at the center of the Coulomb blockade valley ($\epsilon_1=-U/2$). The open squares and filled circles are experimental data,\cite{Liu2009} while the blue (solid) and red (dot-dashed) lines are NRG calculations for $U/\Gamma = 8$, $\lambda = 0.7$, and $U/\Gamma=7.2$, $\lambda = 0.65$, respectively.  The dashed line is $\Delta_G=0$, and the dotted line is $\Delta_G = k_\mathrm{B} T$ with the experimental $T=55$ mK.}
\end{figure}

With the parameters thus chosen, \fref{fig:ExpSplit} compares the size of the peak splittings $\Delta_G$ (defined as half the peak to peak splitting in the finite-bias conductance, using the notation of ref.~\onlinecite{Liu2009}) from theory -- using the approximate \eref{eqn:GcVsd} -- and experiment;\cite{Liu2009} both obtained in the centre of the Coulomb valley ($\epsilon_1=-U/2$). 
We have plotted the data in the form employed in ref.~\onlinecite{Liu2009}, subtracting the Zeeman splitting $g\mub B = 2h$ from the actual splitting $\Delta_G$ to emphasize the deviation of the two. The open squares and filled circles are the experimental data for dots I and II respectively (as in Fig. 4 of ref.~\onlinecite{Liu2009}), while the blue (solid) and red (dot-dashed) lines are the corresponding theoretical results. The latter have been obtained at $T=0$: we note that the experimental splittings are generally somewhat in excess of $T$ (see the dotted line in \fref{fig:ExpSplit}), and hence temperature does not play an important role in the analysis. Note also that for fields very close to the coalescence point where $\Delta_G$ vanishes (i.e. when the splittings approach the dashed curve  in \fref{fig:ExpSplit}) it is difficult to extract the precise value of the splitting, and hence we show only the sections of the curves for which the splitting can be determined reliably.

The agreement between theory and experiment is very good. In both cases theory reproduces well the low-field splittings, and the curves track the experimental results up to fields of around $5\text{T}$, corresponding to Kondo peak splittings of around $1$--$2 \tkn$. At higher fields the theoretical curves certainly deviate from experiment, which we take to be a sign of the breakdown of the quasi-equilibrium approximation in \eref{eqn:GcVsd}. Recently Schmitt and Anders have extended their non-equilibrum scattering-states NRG approach to the Anderson model in a magnetic field;\cite{Schmitt2011} this approach offers a promising means of determining peak splittings out of equilibrium, and further work comparing its predictions with those of \eref{eqn:GcVsd} should help to establish the regimes of the model where non-equilibrium effects play a large role. One interesting question to be pursued here is the effect of left-right asymmetry in the coupling to the leads,\cite{Schmitt2011} since within the quasi-equilibrium approximation this affects only the dimensionless $G_{0}$ in \eref{eqn:GcVsd} and thus simply rescales the conductance uniformly.

The final point to note here is that theory reproduces the crossing of the two curves identified in the experiment. We find this to be entirely a consequence of the slightly different $\lambda$s for dots I and II: if one repeats the calculations with equal $\lambda$s, the curves do not cross. This in fact is a special case of a
more general finding: for a given $\lambda$, our calculations show that curves with different $U/\Gamma$ never cross (even if one or both curves correspond to non-universal parameter regimes). Hence the crossing of the two curves should not be taken\cite{Liu2009} to indicate the breakdown of universal scaling \emph{per se}.

\subsection{Carbon nanotube quantum dots}
\label{ssec:transition}
We now turn to an analysis of the experiments of Quay et al.,\cite{Quay2007} in which the magnetic field dependences of both spin-$\half$ and spin-$1$ Kondo effects were measured in different Coulomb valleys of a carbon nanotube device. 

\subsubsection{Spin-$\half$ Kondo valley}
In the spin-$\half$ valley, conductance maps were obtained\cite{Quay2007} at zero and finite-$B$ as a function of gate and source-drain biases, and the evolution of the finite-bias conductance was also measured as a function of $B$. The splitting of the Kondo resonance at finite bias was compared to various theoretical predictions in the literature, the level of agreement being rather poor.\cite{Quay2007} 
In this section we explain why the experiment did not recover the expected behavior. First, we again parameterize the model from zero-field experimental data.

As before, the spin-$\half$ Kondo effect in experiment is captured well by the Anderson impurity model, \eref{eqn:AIM}. By comparing to the experimental conductance maps in ref.~\onlinecite{Quay2007} we find the value $U/\Gamma = 8.5$ gives optimal agreement with the experimental data at both zero and finite fields. The value of $\lambda\simeq 0.58$ can separately be extracted as described in the previous section,\cite{JCPCNT2009} 
and the experimental $U$ can be determined from the Coulomb peak position in Fig. 2(d) of ref.~\onlinecite{Quay2007}: it is readily seen to be approximately 
$2\text{ meV}$, and hence $\Gamma \simeq 0.24\text{ meV}$. From NRG calculations at $T=0$ we then find $\tilde{T}^{0}_{K}=0.094\Gamma \simeq 0.022\text{ meV} \simeq 0.25\text{ K}$, lower than the experimentally estimated value of $2\text{ K}$. This means that the temperature of the device 
($T=352\text{ mK}$)\cite{Quay2007} is then on the order of the Kondo scale, rather than being somewhat less than it. We believe this to be more consistent with experiment, as now explained.

The magnitude of the experimental Kondo scale can be gauged by inspection of Fig.~2(d) of ref.~\onlinecite{Quay2007}. If these results were obtained at a temperature somewhat below $\tilde{T}^{0}_{K}$, the Kondo resonance would hardly be eroded by temperature, and instead one would naturally attribute the diminution of the zero-bias conductance from the unitarity limit of $2e^2/h$ to the asymmetry of the left and right dot-lead couplings (manifest in  a $G_0\approx 0.4$). But it is then difficult to explain the heights of the Kondo resonance \emph{relative} to that of the Coulomb peaks since (with the caveat that \eref{eqn:GcVsd} is approximate) we would expect\cite{Logan1998} the latter to be around a quarter of the height of the former for $T\ll \tilde{T}^{0}_{K}$. 
We believe it much more likely that $\tilde{T}^{0}_{K}$ is closer to the temperature of the device, eroding more the Kondo resonance and thus reducing its height to something closer to that of the Coulomb peaks.
 
\begin{figure}
\includegraphics{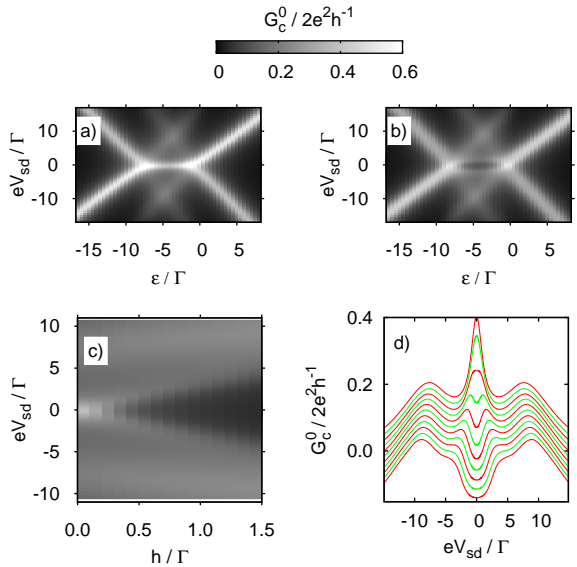}
\caption{\label{fig:AndDC}
Spin-$\half$ conductance maps at zero and finite magnetic field: to compare with fig. 2 of ref. \onlinecite{Quay2007}. An AIM is used with $\util=8.5$ (giving $\omega_K/\Gamma \simeq 0.15$), with $T=\omega_K$ and $G_0=1$. (a) Zero field conductance: a clear Kondo ridge is seen at zero bias voltage. (b) A finite field, $\htil=0.5$, splits this ridge into twin peaks away from zero bias. (c) Evolution of the Kondo peaks with field at the middle of the Coulomb valley (p-h symmetry). (d) Vertical slices through c) with (top to bottom) $\htil=0,0.1,0.2,...,0.9$, offset by $0.02~2e^2h^{-1}$ per slice.
}\end{figure}

Moving on to our conductance results, \Fref{fig:AndDC}(a) shows the theoretical conductance map, calculated from \eref{eqn:GcVsd}, to be compared to Fig.~2(a) of ref.~\onlinecite{Quay2007}. The general agreement is good; the theory reproduces the intense sequential tunneling peaks when the dot level is resonant with one of the lead chemical potentials, the somewhat weaker Coulomb diamond, and the narrower Kondo resonance at zero-bias near the centre of the Coulomb blockade valley. \Fref{fig:AndDC}(b) shows the effect of switching on a magnetic field $h/\Gamma = 0.5$: again, the qualitative agreement with experiment is very good, including now a clear `ellipsoidal' ring around the centre of the Coulomb blockade valley resulting from the splitting of the Kondo resonance by the field.

The field dependence is shown in more detail in \fref{fig:AndDC}(c) and (d), which both show the field-dependence of the conductance in the center of the Coulomb blockade valley. We again recover the key features of the experiment (Fig. 2(c) and (d) of ref.~\onlinecite{Quay2007}). For fields sufficiently small compared to the zero-field Kondo scale ($h/\Gamma \lesssim 0.2$) the Kondo resonance remains intact, while for larger fields it is progressively split and ultimately destroyed with increasing $h$, eventually leading to a region of almost zero conductance around zero bias. Comparing the slices through the data in \fref{fig:AndDC}(d) to those of the experiment, we again observe good qualitative agreement between the two.

We should point out at this stage that the value of $h/\Gamma=0.5$ chosen in \fref{fig:AndDC}(b) corresponds, in physical units, to a field of about $2$ T, around half the experimental field. This again we attribute primarily to the breakdown of the quasi-equilibrium approximation \eref{eqn:GcVsd} at fields larger than a couple of $\tk$: as seen earlier in \fref{fig:ExpSplit} the approximation tends to overestimate the splitting at these high fields, and hence a smaller $h/\Gamma$ must be used in the calculation to obtain the same absolute splitting as the experiment. Based on the comparison of the previous section, noting the theoretical value here of $\tilde{T}^{0}_{K}=0.094\Gamma$, we would estimate that the quasi-equilibrium approximation begins to break down for this experiment at fields $B \gtrsim 1$ T.

While the latter means we cannot compare quantitatively our NRG predictions at finite field to those of the experiment over the whole range of fields measured, we nonetheless believe the parameterization of the experiment to be reliable at low magnetic fields. This allows us to make order-of-magnitude predictions that explain the significance of the experimental results and the reason for the apparent disagreement with theory,\cite{Quay2007}
as now explained.

To summarize the analysis of ref.~\onlinecite{Quay2007}: first the splitting of the Kondo peak with field was extracted from the experimental data and plotted versus $B$. It was found that half the splitting tends to the form $\delta = g\mub B$ at high field (with $g\simeq 2.07$). Direct comparison was then made between the full field-dependence of the splitting obtained from several theories, and experiment.

We point out that there are two basic problems with making this comparison. First and foremost, if one is interested in the \emph{universal} form of the Kondo splitting, the experimental parameters need to satisfy both $U/\Gamma\gg 1$ and $h\ll \min(\Gamma, U)$. The former condition is necessary to ensure that the experiment is well-described by an effective Kondo model at low energies, and arises because the Schrieffer-Wolff transformation that maps the full Anderson model onto the Kondo model is formally valid in the asymptotic limit $U/\Gamma\gg 1$. The latter condition defines what is meant here by `low energies': even if $U/\Gamma$ is large, the effective Kondo description will always break down at energies of the order of the non-universal scale $\Gamma$, and the results on such an energy scale will simply not show universal Kondo form.

One could argue that the $U/\Gamma\simeq 8.5$ here is sufficiently large for the experiment to be well described by a Kondo model at zero field, although we believe this to be a somewhat more borderline case. The main problem however is that the experimental $U$ and $\Gamma$ are too small for the high-field results to be universal. This can in fact be seen directly from Fig.~2(d) of ref.~\onlinecite{Quay2007} (see also \fref{fig:AndDC}(b,d)): even at moderate fields of $2$--$3\text{ T}$ the Kondo (`Zeeman') peaks are already overlapping significantly the non-universal Coulomb peaks. More formally, to be universal for some given $h/\tk$ requires $h/\Gamma \ll 1$; the experimental $U=2\text{ meV}$,\cite{Quay2007} and $\Gamma\simeq 0.2\text{ meV}$ as above, whence $g\mub B\ll \Gamma$ when $B\ll 2\text{ T}$.

The second problem is that the predictions for the theoretical conductance\cite{Logan2001,Costi2000,Moore2000} have all been made using (either explicitly or implicitly) the approximation of \eref{eqn:GcVsd}, rather than from a full-blown non-equilibrium approach. Even when we use the appropriate non-universal parameters in our NRG calculations, the comparison to both the present experiment and that of the previous section suggests that \eref{eqn:GcVsd} is quantitatively reliable only for fields smaller than a few Kondo scales,\cite{Hewson2006} and even then is strongly dependent on the value of $\lambda$. Until non-equilibrium approaches such as the scattering-states NRG\cite{Schmitt2011} become more feasible, the quantitative, universal form of the conductance splitting for $h \gg \tilde{T}^{0}_{K}$ is an open question; one should certainly not expect \emph{a priori} to obtain quantitative agreement between experiment and \eref{eqn:GcVsd}.

\subsubsection{Spin-$1$ Kondo valley}

Finally we consider the effect of magnetic field on the conductance of the two-level model,
\eref{eqn:2levdot}, to make comparison with the spin-$1$ Kondo valley experiments
of ref. \onlinecite{Quay2007}. Since there are more interactions in the two-level Hamiltonian than the AIM, it is obviously harder to parameterize the model from the available data in a fully systematic manner. We have thus endeavoured to choose physically reasonable parameter values  that reproduce qualitatively the experimental results (c.f. those used for the experimental comparison of ref.~\onlinecite{Logan2009}); from which we find $U/\Gamma = 12$, $U'/\Gamma=6$, $J_\mathrm{H}/\Gamma = -0.5$ and $(\epsilon_2-\epsilon_1)/\Gamma\equiv\Delta\epsilon/\Gamma = 4.5$. For simplicity we take $T=0$, $\lambda = \half$ and $G_0=1$. 
Choosing a reasonable value of $\Gamma\sim 0.25\text{ meV}$ gives e.g. a charging energy $U\sim 3\text{ meV}$ 
and level spacing $\Delta\epsilon \sim 1\text{ meV}$, both of which are within typical experimental estimates. 

\begin{figure}
\includegraphics{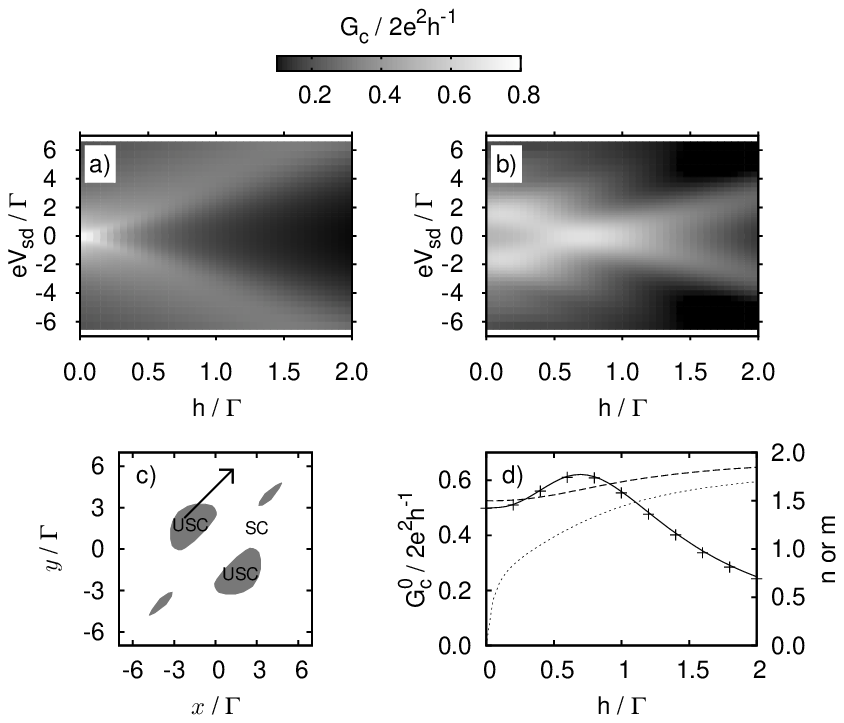}
\caption{\label{fig:TripDC}
Conductance maps for the 2LM close to the $h=0$ phase transition; for $\util=12$, $\util'=6$, $\jtil=-0.5$ and $\Delta\etil=4.5$. To be compared to figs 5a,b of ref. \onlinecite{Quay2007}. \\  {\bf(a)} USC phase for $y=-x = 2.25\Gamma$ ($x=\epsilon_{1}+\half U +U^{\prime}$ and $y=x+\Delta\epsilon$, with $\epsilon_{1} \propto V_{\mathrm{gate}}$). The conductance peak splits on applying a field. {\bf(b)} SC phase for $(x,y) =(1.25\Gamma, 5.75\Gamma)$. 
The $h=0$ Kondo anti-resonance is `filled-in' for $h>0$; full discussion in text.
{\bf(c)} $h=0$ phase diagram for above parameters as a function of $x/\Gamma$ and $y/\Gamma$, with the arrow showing the `trajectory' taken in going from (a) to (b). {\bf(d)} Zero-bias cut through (b) (crosses), along with the total dot occupation number $n(h)$ (dashed), magnetization $m(h)$ (dotted) and $G_{c}^{0}/(2e^2/h)$ as given by \eref{eqn:condA} (solid).
}\end{figure}

\Fref{fig:TripDC}(a) shows the resultant splitting of the `underscreened Kondo' conductance peak at a point in the USC phase near the zero-field  USC/SC phase transition (as indicated by the tail of the arrow in the phase diagram  \fref{fig:TripDC}(c)). The figure is qualitatively similar to that for the spin-$\half$ Kondo peak in a magnetic field [\fref{fig:AndDC}(c)] but, as noted in the case of a stretched spin-$1$ molecule,\cite{Roch2009} the field at which the zero-bias peak is destroyed is a somewhat smaller fraction of the Kondo scale. 
This naturally reflects the sharper USC Kondo resonance (see \fref{fig:TripSplit} left)
compared to the spin-$\half$ Kondo case. 

The Kondo scale $\omega_K^\text{S=1}$ for the chosen parameters is $\omega_K^\text{S=1} \simeq 0.66\Gamma \simeq 0.2\text{ mev}\simeq 2\text{ K}$, which again appears roughly in line with the widths of the Kondo peaks in the experimental conductance maps.\cite{Quay2007} This means it is perhaps misleading to refer to the basic phenomenology here as `underscreened Kondo' physics, since 
resonance widths on the order of $\Gamma$ imply the model is far from being well described by a effective spin-$1$ Kondo model \emph{per se}. As for the semiconducting quantum dot analyzed previously\cite{Logan2009} it also appears that the experimental trajectory as a function of gate voltage ($\epsilon_{1} \propto V_{\mathrm{gate}}$) just cuts the `edge' of the USC phase where the USC-phase Kondo scale is relatively high.

Just across the phase boundary into the SC Fermi liquid phase, we obtain the conductance map shown in \fref{fig:TripDC}(b). Here we have kept the interactions and $\Delta\epsilon$ fixed, but increased $\epsilon_1$ by $3.5\Gamma$ from its value in \fref{fig:TripDC}(a) (the head of the arrow in \fref{fig:TripDC}(c) gives the precise location relative to the phase boundary). The qualitative agreement between theory and experiment (Fig.~5(b) of ref.~\onlinecite{Quay2007}) is again very good. We recover all basic features seen in experiment:\cite{Quay2007} at zero field the conductance peaks at around $\pm 2\Gamma \simeq \pm 0.5\text{ meV}$, reflecting at zero bias the \emph{antiresonance} in the equilibrium spectrum just inside the SC phase.\cite{Logan2009,Hofste2001} These peaks move toward each other, cross and ultimately move apart with increasing field, which can be loosely associated with a crossing of the isolated dot singlet and lowest triplet states, with a finite-field Kondo effect taking place at the crossing point (again, the `Kondo' scale here is rather large, and as such one cannot describe the low-energy behavior in terms of a pure spin-$\half$ Kondo model). We  note that in our calculations the crossing takes place at $h\sim 0.7\Gamma\simeq 0.2\text{ meV}$ and hence $B\sim 3\text{ T}$, again in good agreement with experiment. 
One can also make out various weaker features in the conductance, parallel to the main features and again seen experimentally, which mirror transitions from the isolated dot singlet to the higher energy triplet states.\cite{Quay2007}

The zero-bias conductance is analyzed further in \fref{fig:TripDC}(d), which is a cut through \fref{fig:TripDC}(b) at $eV_\mathrm{sd}=0$ (crosses are the NRG data from \fref{fig:TripDC}(b)).
With increasing field, $G_{c}^{0}/(2e^2/h)$ increases from its zero-field value of $\sim \half$, passes through a maximum at $h/\Gamma \sim 0.7$ (as evident from \fref{fig:TripDC}(b)), and decreases monotonically thereafter.
Also shown are the total dot occupation $n(h)$ and  magnetization $m(h)$ (see \eref{eqn:nmdef}), both of which increase smoothly and monotonically as the ground state evolves with increasing field.
At zero field the dot is in a mixed-valent regime, with $n(h=0)\simeq 1.5$ (and $m=0$). But with increasing  field 
the dot ground state becomes progressively more like the simple $S_z=1$ component of the isolated-dot triplet, with both total charge and magnetization tending to $2$ for $h/\Gamma \gg 1$ (i.e. $n_{\mathrm{imp},\uparrow} \simeq 2$,
$n_{\mathrm{imp},\downarrow} \simeq 0$).

While the zero-bias conductance shown above is calculated using \eref{eqn:MWGc}, and as such probes single-particle spectra, its field-dependence shown in \fref{fig:TripDC}(d) should equally be explicable from \eref{eqn:condA} (\sref{ssec:zerobiascond}), expressed solely in terms of the dot charge and magnetization. That this is indeed so is shown directly in  \fref{fig:TripDC}(d): the solid line is calculated from \eref{eqn:condA}, and seen to be in very good agreement with the direct NRG calculations.

\section{Concluding remarks}
\label{sec:conc}

In this paper we have considered in some detail the effects of an applied magnetic field on single- and two-level quantum dots tunnel-coupled to a metallic lead; including magnetization, single-particle dynamics and conductance, and highlighting for the two-level model in particular the rather subtle differences between the limits $h=0$ and $h\to 0$. We have used NRG to analyze critically the field-dependent shift of the Kondo resonances in the two models, providing an algorithm that in principle can generate the universal scaling behavior over arbitrarily large 
$h/\omega_\mathrm{K}$ ranges, limited in practice only by the logarithmic broadening inherent to NRG. For the single-level AIM, calculations can now be performed sufficiently accurately to achieve convergence up to fields of around $h\sim 100\omega_\mathrm{K}$; for the two-level model convergence is slower, but appears to indicate a constant spectral shift $\Delta_\sigma = 2h$ for all fields in the universal regime.

We have also made direct comparison between NRG calculations and two recent sets of conductance experiments on quantum dots in a magnetic field,\cite{Liu2009,Quay2007} using Anderson-type models for the dots to determine 
bare parameters corresponding to the experimental realizations. Agreement between theory and experiment is found to be very good qualitatively (essentially all salient experimental features are captured by the models), and even quantitatively -- see e.g. \fref{fig:ExpSplit} -- provided the system is not `too far out of equilibrium'. Since NRG provides in essence numerically-exact results, the deviation of calculations from experiment provides a measure
of the quantitative reliability of the quasi-equilibrium approximation in \eref{eqn:GcVsd}, used throughout to calculate conductance; we find it typically breaks down when the field-induced splitting exceeds somewhat the zero-field Kondo scale.

We have argued that neither experiment\cite{Liu2009,Quay2007} considered has measured the \emph{universal} conductance splitting of the spin-$\half$ Kondo effect; and have emphasized (\sref{ssec:liu}) the considerable sensitivity of the field-dependence of the conductance peak to the partitioning of the bias potential between the leads (embodied in $\lambda$) -- over which, to our knowledge, there is relatively little experimental control. In addition, as above, \eref{eqn:GcVsd} for the conductance is approximate out of equilibrium, and until non-equilibrium approaches such as e.g. the scattering states NRG\cite{Schmitt2011} are sufficiently developed to become the mainstay, present theoretical tools are  limited in that respect.

 In view of the above, we suggest that more experimental attention should be given to 
the \emph{equilibrium}, zero-bias conductance. Given exactly by \eref{eqn:MWGc}, and independent of $\lambda$, its field dependence can be calculated exactly (see e.g. \sref{ssec:zerobiascond}).
We believe it presents a better prospect for ascertaining universality in the magnetic field dependence of spin-$\half$ and spin-$1$ Kondo effects in real quantum dots.

\begin{acknowledgments}
We thank the EPSRC-UK for financial support.
\end{acknowledgments}



\begin{thebibliography}{65}
\bibitem[{not({\natexlab{a}})}]{note:tk}
\bibinfo{note}{$\tk$ is the $h=0$ Kondo scale, defined\cite{Logan2009} for the
  AIM by $\Simp(\tk)=0.1$ with $\Simp(T)$ the dot (impurity) contribution to
  the entropy of the system.}
\bibitem[{not({\natexlab{c}})}]{note:bw}
\bibinfo{note}{The bandwidth $D$ is taken to be sufficiently larger than all
  other energies that its precise value is immaterial.}
\end{thebibliography}

\end{document}